\documentclass[final,3p,times,twocolumn]{elsarticle}
\pdfoutput=1

\usepackage{preprintcover} 

\PreprintCoverPaperTitle{A search for prompt lepton-jets in $pp$
  collisions at $\sqrt{s}=7$\,TeV with the ATLAS
  detector} 

\PreprintIdNumber{CERN-PH-EP-2012-319} 

\PreprintCoverAbstract{ We present a search for a light (mass $<$ 2
  GeV) boson predicted by Hidden Valley supersymmetric models that
  decays into a final state consisting of collimated muons or
  electrons, denoted ``lepton-jets''. The analysis uses 5~fb$^{-1}$ of
  $\sqrt{s} = 7$~TeV proton--proton collision data recorded by the
  ATLAS detector at the Large Hadron Collider to search for the
  following signatures: single lepton-jets with at least four muons;
  pairs of lepton-jets, each with two or more muons; and pairs of
  lepton-jets with two or more electrons.  This study finds no
  statistically significant deviation from the Standard Model
  prediction and places 95\% confidence-level exclusion limits on the
  production cross section times branching ratio of light bosons for
  several parameter sets of a Hidden Valley model.}  

\PreprintJournalName{Phys. Lett. B} 

\usepackage{atlasphysics}
\usepackage{graphicx}
\usepackage{dcolumn}
\usepackage{bm}
\usepackage{hyperref}
\usepackage[mathlines]{lineno}
\usepackage[subrefformat=parens,labelformat=parens]{subfig}
\usepackage{amssymb}

\def\alphad{$\alpha_d$}
\def\gammad{$\gamma_D$}
\def\mgammad{$m_{\gamma_{D}}$}
\def\MUONLUMINOSITYNOUNIT{4.5}
\def\MUONLUMINOSITY{\MUONLUMINOSITYNOUNIT~fb$^{-1}$}
\def\ELECTRONLUMINOSITY{4.8~fb$^{-1}$}
\def\fHT{\ensuremath{f_{\mathrm{HT}}}}
\def\fEM{\ensuremath{f_{\mathrm{EM}}}}

\biboptions{sort&compress}

\begin{document}

\title{A search for prompt lepton-jets in $pp$ collisions at $\sqrt{s}=7$\,TeV with the ATLAS detector}

\author{ATLAS Collaboration}

\date{\today}

\begin{abstract}
  We present a search for a light (mass $<$ 2 GeV) boson predicted by
  Hidden Valley supersymmetric models that decays into a final state
  consisting of collimated muons or electrons, denoted
  ``lepton-jets''. The analysis uses 5~fb$^{-1}$ of $\sqrt{s} = 7$~TeV
  proton--proton collision data recorded by the ATLAS detector at the
  Large Hadron Collider to search for the following signatures: single
  lepton-jets with at least four muons; pairs of lepton-jets, each
  with two or more muons; and pairs of lepton-jets with two or more
  electrons.  This study finds no statistically significant deviation
  from the Standard Model prediction and places 95\% confidence-level
  exclusion limits on the production cross section times branching
  ratio of light bosons for several parameter sets of a Hidden Valley
  model.
\end{abstract}

\maketitle

\section{Introduction}

A light boson at the GeV scale, in a model where a Hidden Valley sector is weakly
coupled to the Standard Model
(SM) sector~\cite{DMtheory1,DMtheory2,Strassler}, 
has been proposed to explain several recently observed
anomalies in cosmic-ray and dark matter direct-detection experiments.
These observations include an unexpected excess of cosmic electrons
and/or positrons~\cite{Barwick2009Measurements,FermiLat, PAMELA, ATIC}
and signals from certain dark matter direct-detection
experiments~\cite{DAMA, COGENT, CRESST}.  The proposed boson could be
created at particle accelerators and produce distinctive final states
of tightly collimated ``lepton-jets'' consisting of close by 
electrons or muons~\cite{Yavin1, wacker, kribs,
sundrum, falkowski}.  Such lepton-jet decays are also a generically
interesting signature that may be produced by rare decays of, for
instance, $Z$ or Higgs bosons~\cite{Yavin2}.  Upper limits on lepton-jet production
have already been set by previous analyses of collider data~\cite{cms_lj,
d0_lj}.

In Hidden Valley models, the universe consists of SM and
supersymmetric (SUSY) particles, together with an additional spectrum
of dark matter particles charged under a hidden gauge group (called
the dark sector).  Certain particles called messengers are charged
under both the dark sector and the SM and SUSY gauge symmetries,
permitting decay chains through the normal and dark sectors.
For example, the lightest supersymmetric particle, which cannot
decay to SM particles due to R-parity conservation, can 
decay into less-massive dark-sector states ending with the lightest
particle in the dark sector, a dark photon denoted \gammad. This dark
photon can decay into light SM fermions by kinetic
mixing~\cite{Holdom} of the dark gauge sector and SM gauge
symmetries. These models aim to explain the excess of cosmic-ray
positrons, in the absence of any observed proton excess, with a
dark boson \gammad\ that has a mass below the proton--antiproton
kinematic threshold of $\sim$2 GeV. 
Such low-mass dark photons can decay to electrons, muons, and pions,
whereas decays to protons are kinematically forbidden.
Due to the boost of the \gammad, the light SM decay products are
highly collimated, providing a striking signature for new physics.

The data is interpreted in a model where a pair of squarks is produced
and each of the squarks cascade decays into dark-sector particles,
including one or more dark-photons. The dark-photons decay into pairs
of leptons, forming lepton-jets.  
Additionally, dark-sector
particles may radiate multiple dark photons, increasing the lepton
multiplicities and number of the lepton-jets~\cite{Yavin2}.  The amount of
radiation is determined by the dark sector gauge coupling parameter
\alphad. Setting \alphad\ = 0.0 results in a simple lepton-jet with
two hard leptons. Larger values of \alphad\ may produce lepton-jets
with four, six, eight, or more prompt leptons from the decay of
overlapping dark photons, albeit with reduced boost. The transverse
momentum (\pt) of the leptons increases with dark photon mass, but
decreases with \alphad.  This paper considers values of
\alphad\ of ~0.0, 0.1,
and 0.3, and dark photon masses (\mgammad) of ~150, 300, and 500~MeV.
For \mgammad\ = 150~MeV, the dark photon is below the muon--antimuon
threshold and can only decay to electrons.  With \mgammad\ $\ge
300$~MeV, the dark photon decays to electron and muon
pairs. Additionally, for \mgammad\ = 500~MeV, 20\% of the decays
produce pion pairs. These nine signal operating points cover a wide
range of phase space from low-multiplicity lepton-jets containing
leptons of only one flavour, to high-multiplicity lepton-jets
containing a mix of electrons and muons.

The data samples used in this analysis were collected with the ATLAS
detector during the 2011 run of the Large Hadron Collider at
centre-of-mass energy $\rts = 7$~TeV and correspond to
\MUONLUMINOSITY\ of integrated luminosity for the muon analyses and
\ELECTRONLUMINOSITY\ for the electron analysis~\cite{lumi,lumi2}, after their respective
data quality requirements have been applied.  This paper considers
lepton-jets in three signatures: single muon-jets with four or more
muons, pairs of muon-jets each with two or more muons, and pairs of
electron-jets each with two or more electrons.  The selection is
designed to enhance the signal relative to the SM backgrounds, the
largest of which is multi-jet production. In multi-jet production the
background arises from either real leptons from the decay of SM
particles, from hadrons that are misidentified as leptons, or in the
case of electrons, from photon conversions. All other SM background
sources are expected to be negligible after the final selection cuts
are applied. 
The multi-jet background is reduced through a variety of selection
cuts, and the remaining background is estimated with two different
data-driven techniques.

No requirements are made on the remaining activity in the event beyond
the one or two lepton-jets in order to avoid introducing a strong
model dependence in the analysis. For example, no cuts are made on the
presence of other particles or jets, the event track multiplicity, or
the presence of missing transverse energy.

\section{The ATLAS detector}

ATLAS is a general purpose detector~\cite{atlas} consisting of an
inner tracking detector (ID) embedded in a 2 T solenoid,
electromagnetic and hadronic calorimeters and a muon spectrometer (MS)
employing toroidal magnets. The ID provides precision tracking of charged
particles for $|\eta|<2.5$ using silicon pixel and
microstrip detectors and a straw-tube transition radiation tracker (TRT)
that relies on transition radiation to distinguish electrons from
pions in the range $|\eta| < 2.0$.
Liquid-argon (LAr) electromagnetic sampling calorimeters, with
excellent energy and position resolution, cover the range $|\eta| <
3.2$ with a typical granularity of $\Delta\eta \times \Delta\phi$ of
$0.025 \times 0.025$.  A scintillator-tile calorimeter, which is
divided into a large barrel and two smaller extended-barrel cylinders,
one on each side of the central barrel, provides hadronic calorimetry
in the range $|\eta| < 1.7$. In the end-caps ($|\eta| > 1.5$), LAr
is also used for the hadronic calorimeters, matching the
outer $|\eta|$ limit of end-cap electromagnetic calorimeters. The LAr
forward calorimeters provide both electromagnetic and hadronic energy
measurements, and extend the coverage to $|\eta| = 4.9$.  The
calorimeter system has a minimum depth of 9.7 interaction lengths at
$\eta = 0$. The MS covers $|\eta|<2.7$ and provides triggering and
precision tracking for muons.\footnote{ATLAS uses
  a right-handed coordinate system with its origin at the nominal
  interaction point (IP) in the centre of the detector and the
  $z$-axis along the beam pipe. The $x$-axis points from the IP to the
  centre of the LHC ring, and the $y$ axis points upward. Cylindrical
  coordinates $(r,\phi)$ are used in the transverse plane, $\phi$
  being the azimuthal angle around the beam pipe. The pseudorapidity
  is defined in terms of the polar angle $\theta$ as
  $\eta=-\ln\tan(\theta/2)$.}

A three-level trigger system is used to select events. The
Level 1 (L1) trigger is implemented in hardware and uses information
from the calorimeters and muon sub-detectors to reduce the event rate
to a design value of at most 75~kHz. This is followed by two
software-based trigger levels, Level 2 (L2) and Event Filter (EF),
which together reduce the event rate to 300~Hz on average. The L1 trigger
generates a list of Regions of Interest (RoI) $\eta$--$\phi$ coordinates
with associated thresholds. The muon RoI have a spatial extent of 0.2
in $\Delta \eta$ and $\Delta \phi$ in the MS barrel, and 0.1 in the MS
endcap. The electromagnetic calorimeter RoI have a spatial extent of
0.2 in $\Delta \eta$ and $\Delta \phi$. At L2, most reconstruction
uses simplified algorithms running on data localized to an RoI which
was reported by L1. At the EF level, the trigger system has access to the full event
for processing.

\section{Event reconstruction and selection}

The analysis used only data from stable running periods, and required
events to have a primary collision vertex containing at least three
tracks with $\pt > 400$~MeV in order to remove cosmic rays.

\subsection{Electron-jet channel}

Events containing electron-jets were selected using single-electron
triggers with an online \pt\ threshold of 20 or 22 GeV, the latter
being used after there was a substantial increase in the instantaneous
luminosity during 2011. To ensure
proper modelling of the trigger acceptance, events
were required to contain at least one reconstructed electron with $\pt
> 35$ GeV, above which the trigger efficiency is constant. The
reconstructed electron was required to match an electron reconstructed
above the \pt\ threshold in the trigger system with a separation in
$R$ ($\Delta R \equiv
\sqrt{(\Delta \phi)^2 + (\Delta \eta)^2}$) less than 0.2.

The electron-jet candidates were built from electromagnetic clusters
with minimum transverse energy $\et > 10$~GeV inside the calorimeter
fiducial region ($|\eta| < 2.47$, excluding the 
barrel/end-cap transition region $1.37 < |\eta| <
1.52$ where there is substantial dead material that is difficult to
model accurately). At least two tracks from the primary vertex
(transverse impact parameter
$|d_0| < 1$~mm) having $\pt > 10$ GeV were required to have $\Delta R
< 0.1$ of the cluster position in the second sampling layer of the
calorimeter. Additional requirements were made on the number of hits
along the track in the silicon pixel and silicon microstrip detectors to
suppress backgrounds from photon conversions. The analysis required
two lepton-jet candidates in each event, with one
cluster matching the electron reconstructed in the trigger system. The
invariant mass of the two highest-\pt\ tracks associated with each electron-jet
had to be less than 2 GeV.

\begin{figure*}[ht!]
  \centering
  \subfloat[][]{
    \includegraphics[width=2.1in]{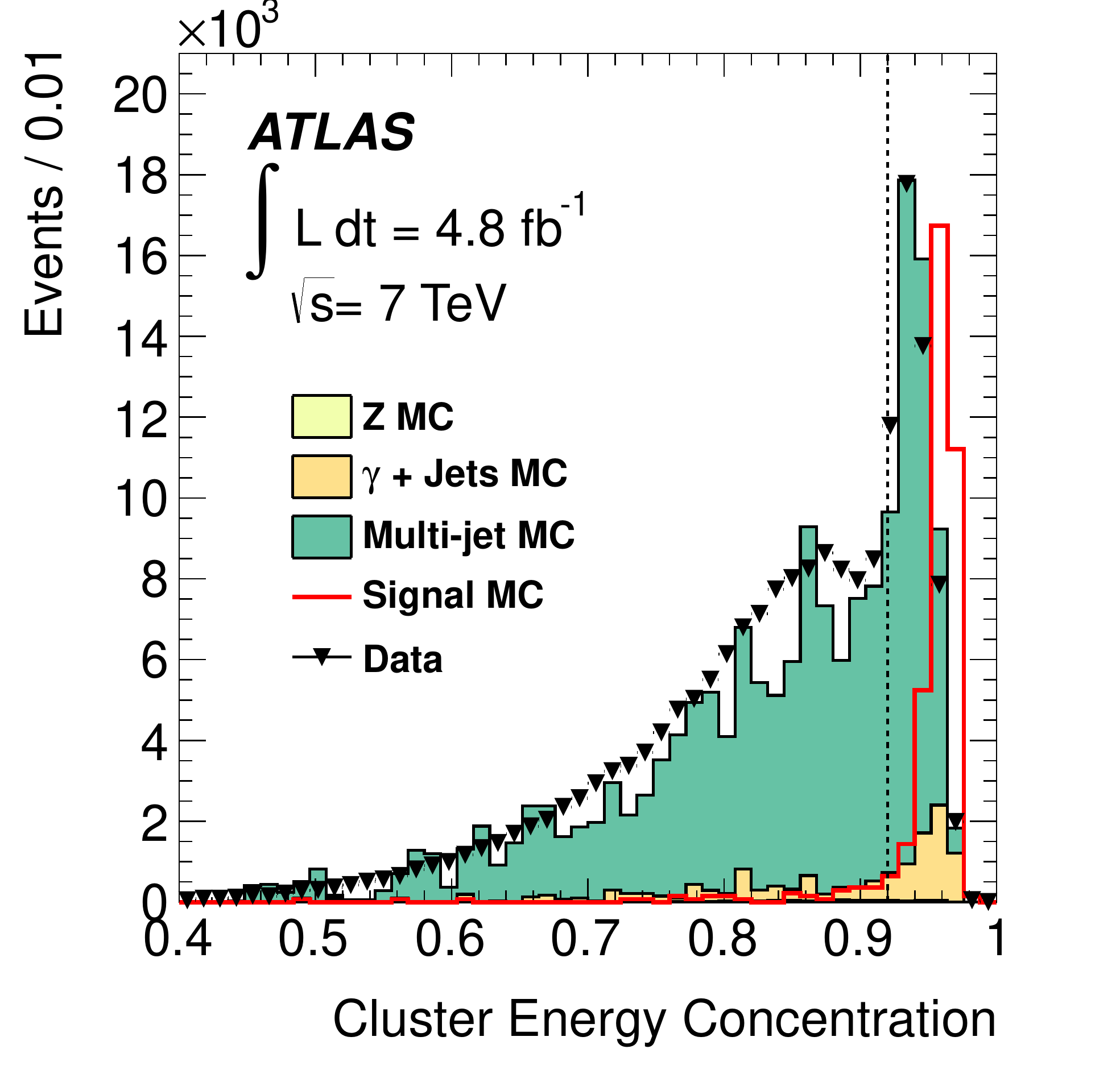}
    \label{fig:discriminating-reta2}
  }
  \subfloat[][]{
    \includegraphics[width=2.1in]{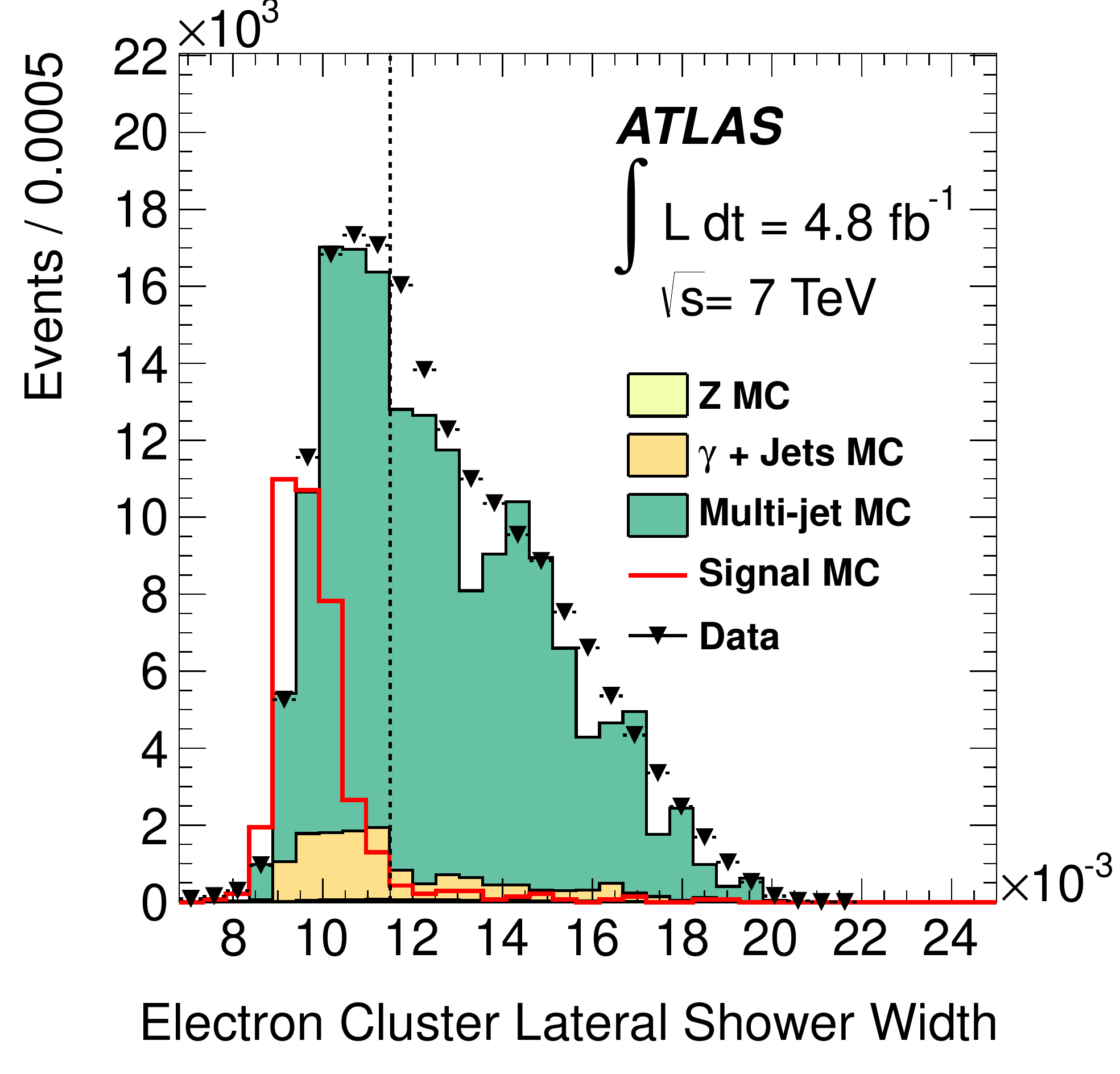}
    \label{fig:discriminating-weta2}
  }
  \subfloat[][]{
    \includegraphics[width=2.1in]{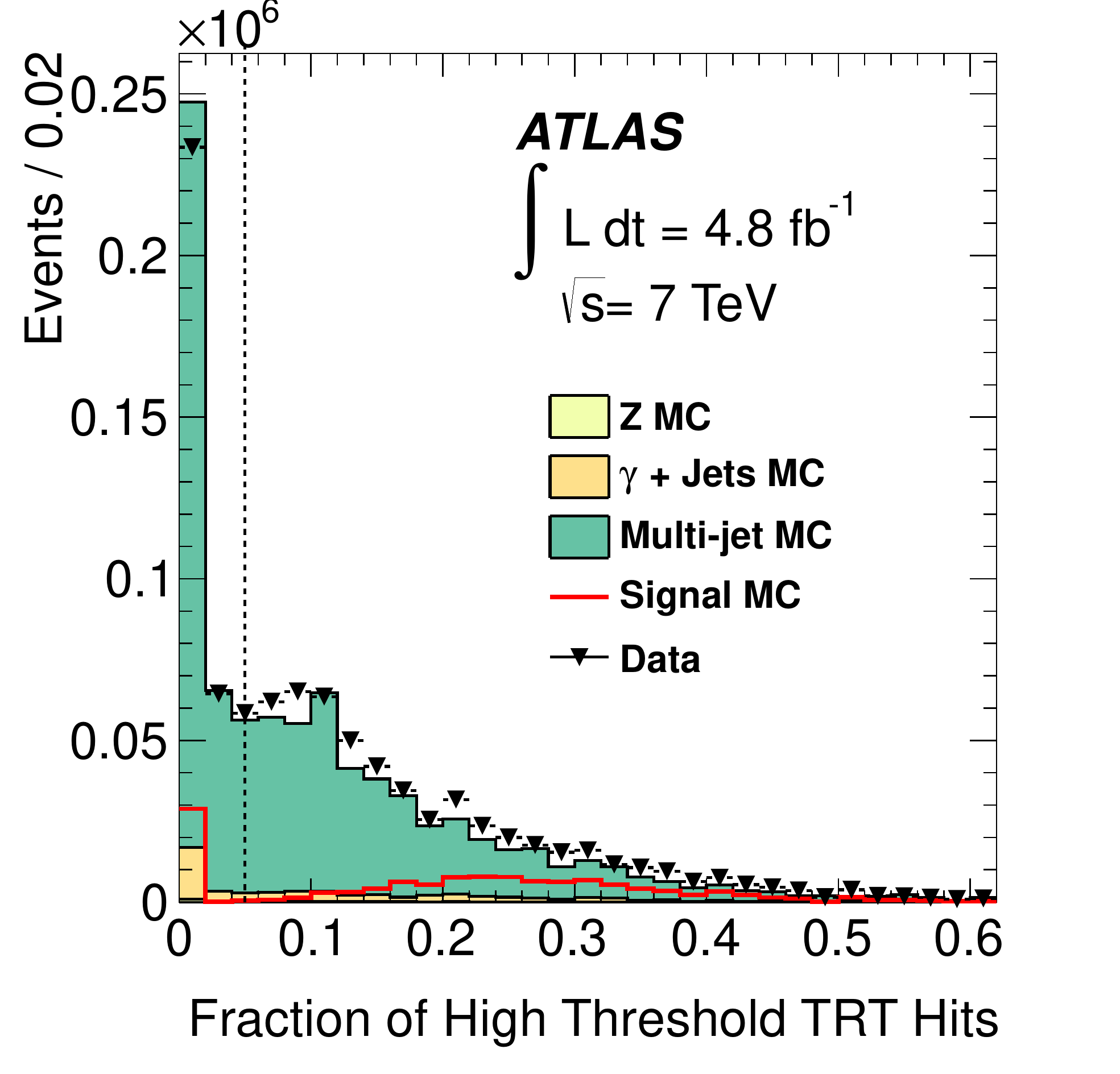}
    \label{fig:discriminating-htfrac}
  }
  \\
  \subfloat[][]{
    \includegraphics[width=2.1in]{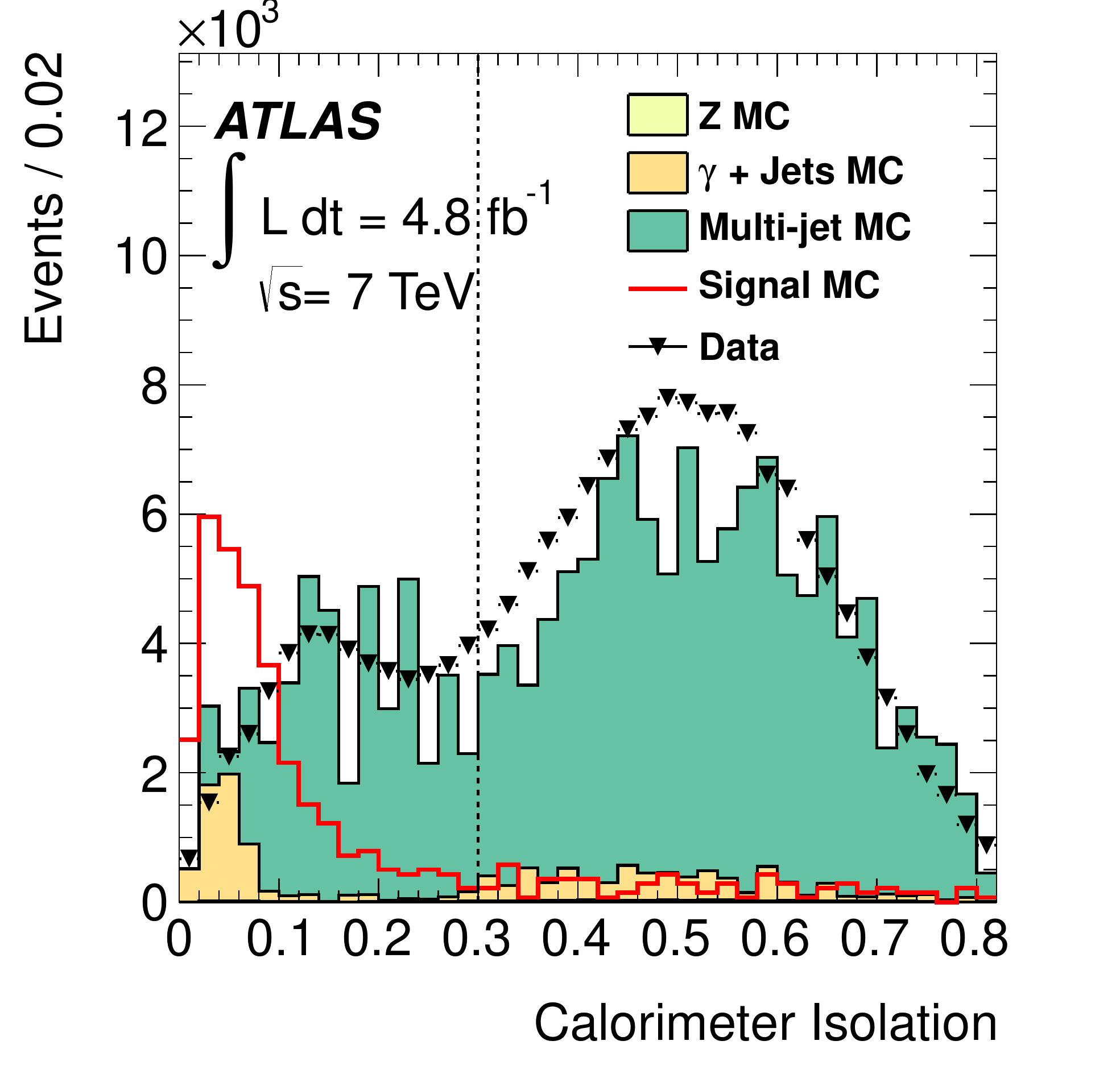}
    \label{fig:discriminating-iso}
  }
  \subfloat[][]{
    \includegraphics[width=2.1in]{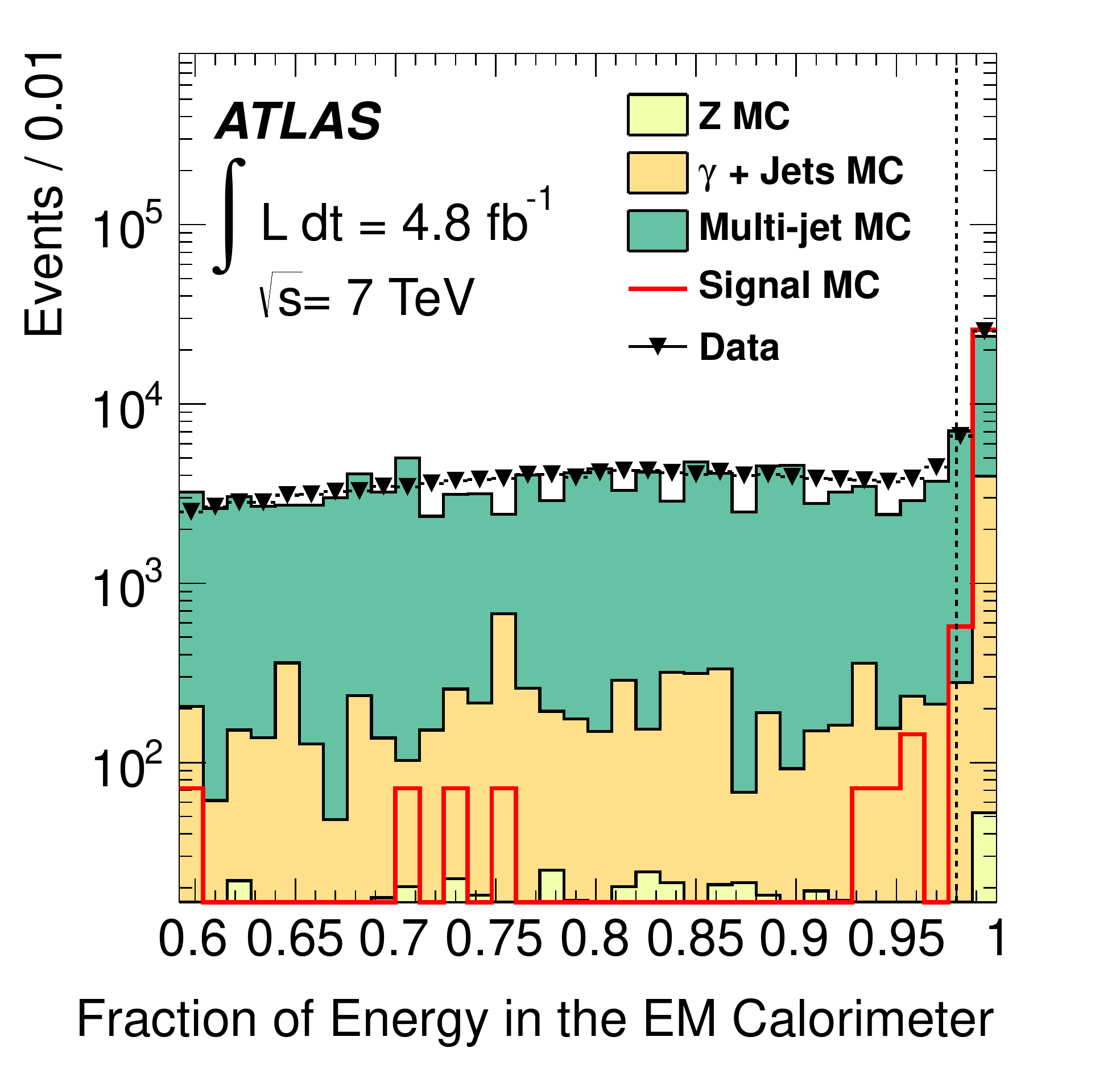}
    \label{fig:discriminating-emfrac}
  }
  \caption{Distributions of the five discriminating
    variables used after selection of events which have passed the
    trigger and contain two or more electron-jet candidates, shown
    separately for the multi-jet, $Z \to ee$ and $\gamma$ + jets
    backgrounds as well as for the signal sample with \alphad\ = 0.0
    and \mgammad\ = 150~MeV. The signal MC normalization is arbitrary. The dashed black vertical line shows the
    cut values at \protect\subref{fig:discriminating-reta2} cluster
    energy concentration $R_{\eta 2} \geq 0.92$,
    \protect\subref{fig:discriminating-weta2} electron cluster lateral
    shower width $w_{\eta 2} \leq 0.0115$,
    \protect\subref{fig:discriminating-htfrac} fraction of high
    threshold TRT hits $\fHT \geq 0.05$,
    \protect\subref{fig:discriminating-iso} calorimeter isolation
    $\leq$ 0.3, and \protect\subref{fig:discriminating-emfrac}
    fraction of energy in the EM calorimeter $\fEM \geq 0.98$. The
    hadronic jet and $\gamma$ + jets distributions are shown here from
    \textsc{Pythia} MC for illustrative purposes. 
    \label{fig:discriminating-variables}
  }
\end{figure*}

The background for the electron-jets analysis comes primarily from
multi-jet events, and to a lesser extent from photon+jet events. Five
variables were used to reduce the remaining background for
electron-jet candidates. The electron cluster energy concentration
shown in Fig.~\subref*{fig:discriminating-reta2}, $R_{\eta 2}$, must
exceed 0.92. $R_{\eta 2}$ is defined as the ratio of total energy
in $3\times 7$ cells to the total energy in $7\times 7$ cells in
$\eta$--$\phi$ in the second sampling layer of the electromagnetic
calorimeter. The electron cluster lateral shower width in the
calorimeter, $w_{\eta 2}$, shown in
Fig.~\subref*{fig:discriminating-weta2}, must be less than 0.0115, where
\begin{equation}w_{\eta 2 } = \sqrt{\frac{\sum_{i} E_i \times
      {\eta_i}^2}{\sum_{i} E_i}-\left(\frac{\sum_{i} E_i \times
        \eta_i}{\sum_i E_i} \right)^2}.\end{equation} Here $E_i$ and
$\eta_i$ represent the energy and pseudorapidity of the $i^{\it{th}}$
cell in a $3\times 5$ $\eta - \phi$ window in the second sampling
layer of the electromagnetic calorimeter. The ratio of the number of
high-threshold hits~\cite{atlas}, indicative of transition radiation, to the total
number of hits from the TRT associated with each track, $\fHT$, was
required to be greater than 0.05 to remove pions. The $\fHT$ distribution per
track is shown in Fig.~\subref*{fig:discriminating-htfrac}. The sharp
peak at zero arises from tracks matched to an electron candidate
outside of the TRT acceptance. A scaled
isolation variable is defined as the transverse energy within $0.1
<\Delta R < 0.4$ around the cluster divided by cluster $\et$; events
were required to have scaled isolation below 30\% as shown in
Fig.~\subref*{fig:discriminating-iso}. Finally, a requirement that the
fraction of the lepton-jet energy found in the electromagnetic
calorimeter, $\fEM$ must be larger than 0.98, was used to reject
activity from hadrons, as shown in
Fig.~\subref*{fig:discriminating-emfrac}.

\subsection{Muon-jet channels}

Single muon-jet events were selected from events satisfying a trigger
with a single muon having more than 18~GeV in \pt. Candidates for
double muon-jets were taken with either a single-muon trigger with a
\pt\ threshold of 18 GeV or a three-muon trigger with a \pT\ threshold
of 6 GeV. The muon triggers were complementary, as the three-muon
trigger has reduced efficiency for high-\pt\ muons from a single
lepton-jet which may be too close together to produce more than one
RoI.

Muon candidates must have been reconstructed in both the ID and the MS
and have $|\eta| < 2.5$. Additional requirements were made on the
number of associated hits in the silicon pixel and microstrip
detectors, as well as on the number of track segments in the MS. The
muons were required to come from the primary vertex by imposing a $|d_0|
< 1$~mm cut on the tracks. The muon-jets were reconstructed in an
iterative procedure using all candidate muons, by seeding the jet
candidate with the highest-\pt\ muon, and adding all muons within
$\Delta R = 0.1$.  Additional jets were formed using the remaining
muons, again seeding the muon-jet with the remaining highest-\pt\
muon.  For the double muon-jet analysis, two muons with $\pt>11$ GeV
were required per jet with the additional requirement that the leading
muon \pt\ be greater than 23 GeV for the single muon trigger events.
For the single muon-jet analysis, four muons were required per jet
with $\pt>$ 19, 16, 14~GeV, respectively, for the three highest-\pt\ muons, and $\pt
> 4$~GeV for all additional muons.

Within a muon-jet, the two muons closest in \pt\ were required to
have an invariant mass less than 2~GeV.  A scaled isolation variable
was formed by summing the \et\ of all calorimeter cells within $\Delta
R = 0.3$ of any of the muon-jet's component muons while excluding
cells found within $\Delta R = 0.05$ of the muons, and dividing by the
muon-jet \pt. The scaled isolation was required to be less than 0.3
(0.15) per muon-jet for the double (single) muon-jet analyses, to
suppress muons from hadronic jets.

\begin{figure}[htb]
  
  \centerline{\includegraphics[width=3.in]{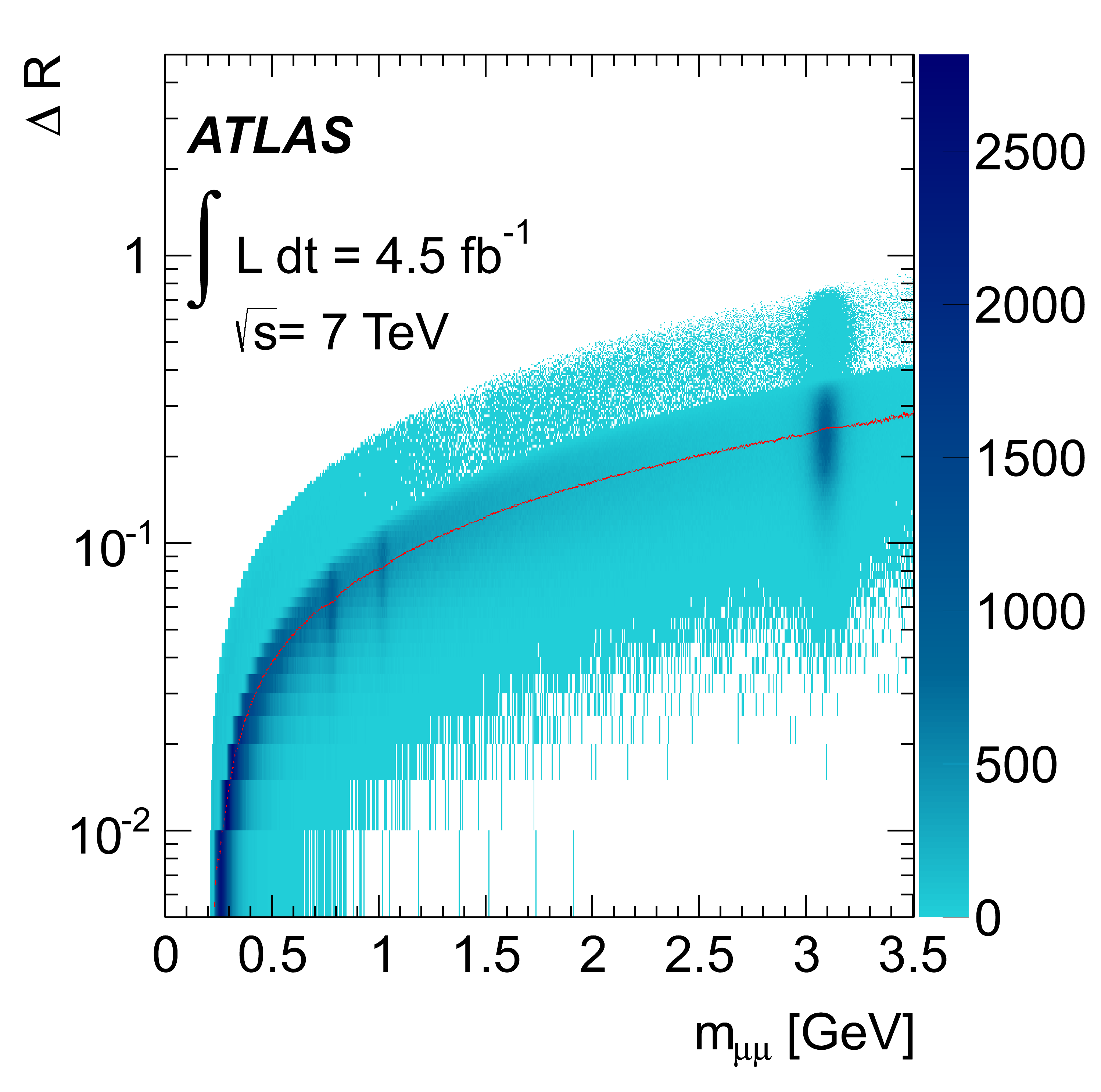}}
  \caption{The opening angle $\Delta R$ vs invariant mass for all muon
    pairs in the 4.5 $\rm fb^{-1}$ data sample. The overlaid red
    points show the profile of the $\Delta R$
    distribution for each dimuon mass
    bin. The position of each point is the mean value of the
    vertical slice and its width is the RMS. The secondary distribution
    running along the top of the distribution arises from
    events with more than two muons, where a third muon triggers at a
    higher \pt, allowing for combinations of dimuon pairs with a
    larger opening angle.
  }\label{fig:dimuoninvmass}
\end{figure}

As noted earlier, a signature of the dark matter signal is a 
muon-jet composed of two or more muon tracks confined to a narrow
cone. One source of collimated muons arises from the decay of low-mass
states, since the opening angle is in inverse proportion to the Lorentz
boost. The background from boosted low-mass states with an invariant
mass less than 3.5 GeV is displayed in Fig.~\ref{fig:dimuoninvmass}
showing the opening angle $\Delta R$ vs invariant mass for all dimuon
pairs. This plot was produced using the same muon selection used for
the muon-jet analysis, excluding the $\Delta R$ requirement. The
invariant mass of muon pairs falls off smoothly, interrupted by easily
observable narrow peaks produced by low-mass resonances such as $\phi$
($\sim$1 GeV), and $\omega$ and $\rho$ ($\sim$0.7 GeV).  For smaller
opening angles, $\Delta R \lesssim 0.03$, the low-mass resonances
barely stand out from the rest of the background. It was not practical
to exclude the $\omega$/$\rho$ and $\phi$ peak regions from the
analysis. However, the $J/\psi$ was removed for the electron-jet and
muon-jet search by a 2 GeV mass cut. A second smoothly falling
distribution is also visible in this figure from events with
an additional three or more muons, one of which has a high enough \pt\
to fire the trigger, producing an additional combinatorial
background.

\section{Signal and background estimation}

Both MC and data-driven methods were used for background and
efficiency estimations.  Various SM processes can mimic the signal due
to misreconstructed objects, such as jets misidentified as electrons,
or chance overlap of leptons. We have considered MC hadronic multi-jet
events, $\gamma$ + jets events, $W \rightarrow \ell\nu +$ jets, $Z
\rightarrow \ell^{+}\ell^{-} +$ jets, $t\bar{t}$ and diboson
($WW$, $WZ$, $ZZ$)  events at $\rts = 7$
TeV. \textsc{Pythia}6~\cite{Pythia} was used for all
samples except $t\bar{t}$, $WW$, $WZ$, $ZZ$ for which
MC@NLO~\cite{MCatNLO} was used. The contribution from $WZ$ and $ZZ$
backgrounds, when one of the bosons is off-shell, was modeled with Sherpa~\cite{Sherpa}.
Of all the backgrounds considered, only the hadronic multi-jet and $\gamma$
+ jets events contribute significantly to the final background
expectation. In addition, signal MC simulation was generated using
\textsc{MadGraph}~\cite{Madgraph} with the CTEQ6L1 set of parton
distribution functions~\cite{MadgraphPDF}, and a custom-made
Mathematica~\cite{Mathematica} 
package to model the dark-sector cascade decay described
in Refs.~\cite{Yavin1, Yavin2}, followed by \textsc{Pythia}6 for hadronization. All
MC samples include the effect of multiple $pp$ interactions per bunch
crossing and are assigned an event weight such that the distribution
of the number of $pp$ interactions matches that in data. The mean
momentum of the dark photons depends strongly on \alphad\ and 
therefore the acceptance of the lepton-jets also depends on this
parameter. At \alphad\ = 0.0 the mean momentum of the dark photon is
73, 76, and 82~GeV for \mgammad\ = 150, 300, and 500~MeV,
respectively, with no cuts applied. For \alphad\ = 0.1, the mean dark
photon momentum decreases to 30.4, 35.9, and 41.6~GeV. At \alphad\ =
0.3 the mean values are 21.1, 25.7, and 30.9~GeV.  All MC events were
processed with the \textsc{Geant4} based ATLAS detector
simulation~\cite{Geant1,atlas_simulation} and then analyzed with the
standard ATLAS reconstruction software.

Due to the very small acceptance for hadronic jets passing our signal
criteria, $\mathcal{O}(10^{-3})$ to $\mathcal{O}(10^{-4})$ for jets
with $50 < \pt < 400$~GeV, there were too few MC events to
accurately estimate background yields. The background MC samples
were used to help
establish the event selection criteria, based on characteristics of
the background.  All the samples were required to satisfy the trigger
conditions, with efficiencies ranging from 40\% to 75\% for the
lepton-jet models considered.

\subsection{Background estimation with the ABCD-likelihood method}
\label{sec:bkg_est}

\begin{figure*}[ht!]
  \centering
  \subfloat[][]{
  \includegraphics[width=2.05in]{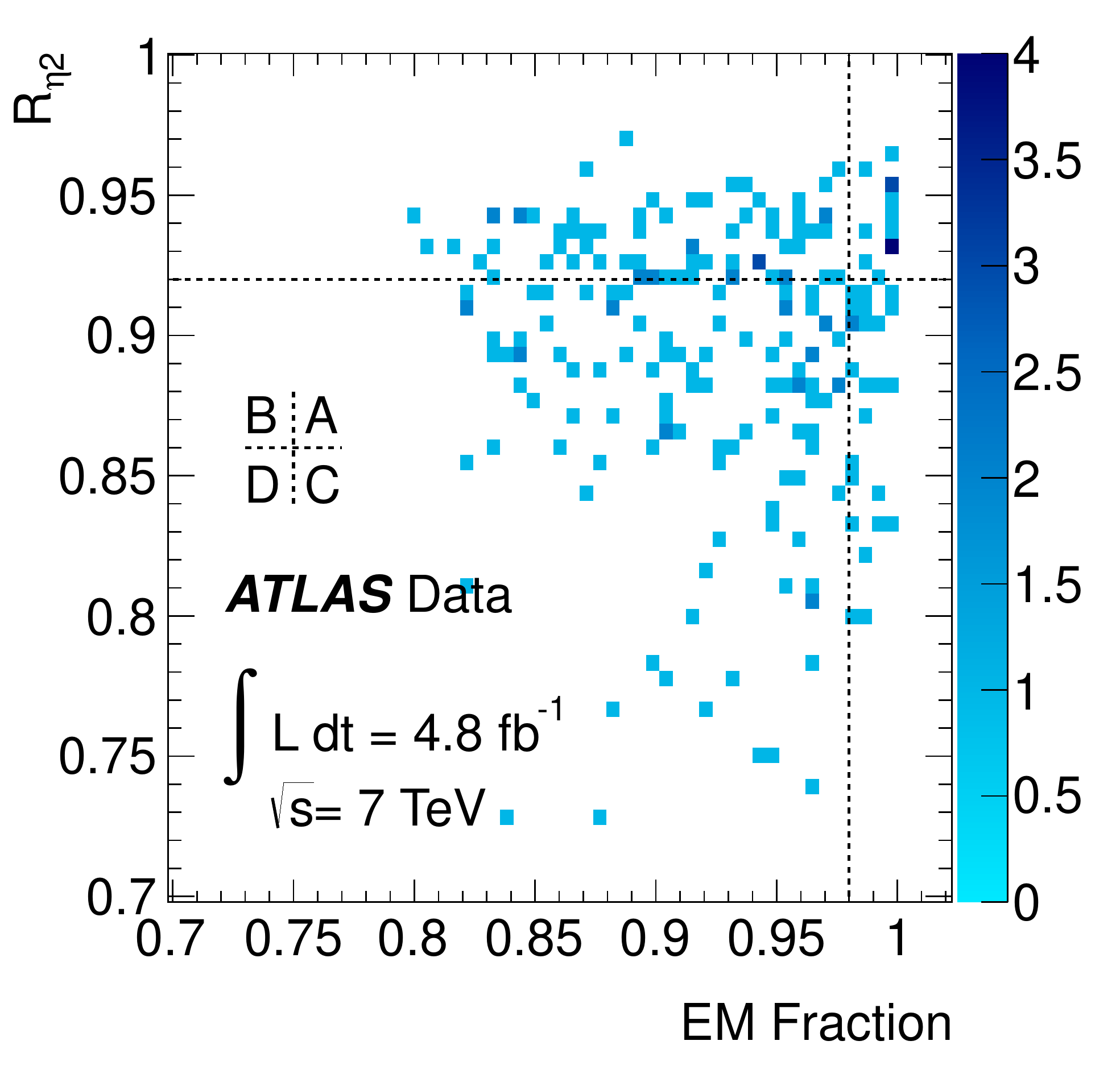}
  }
  \subfloat[][]{
  \includegraphics[width=2.05in]{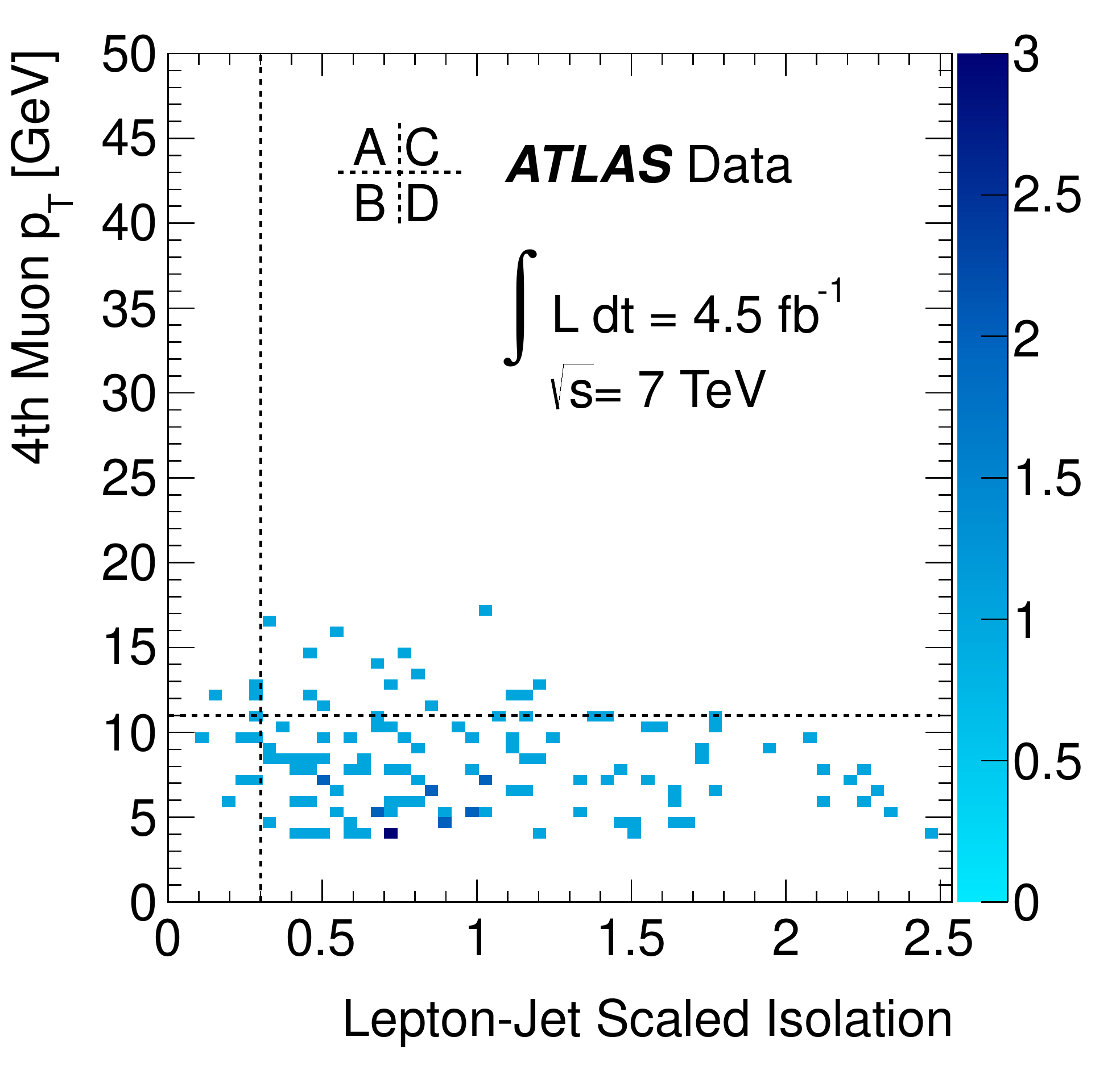}
  }
  \subfloat[][]{
  \includegraphics[width=2.05in]{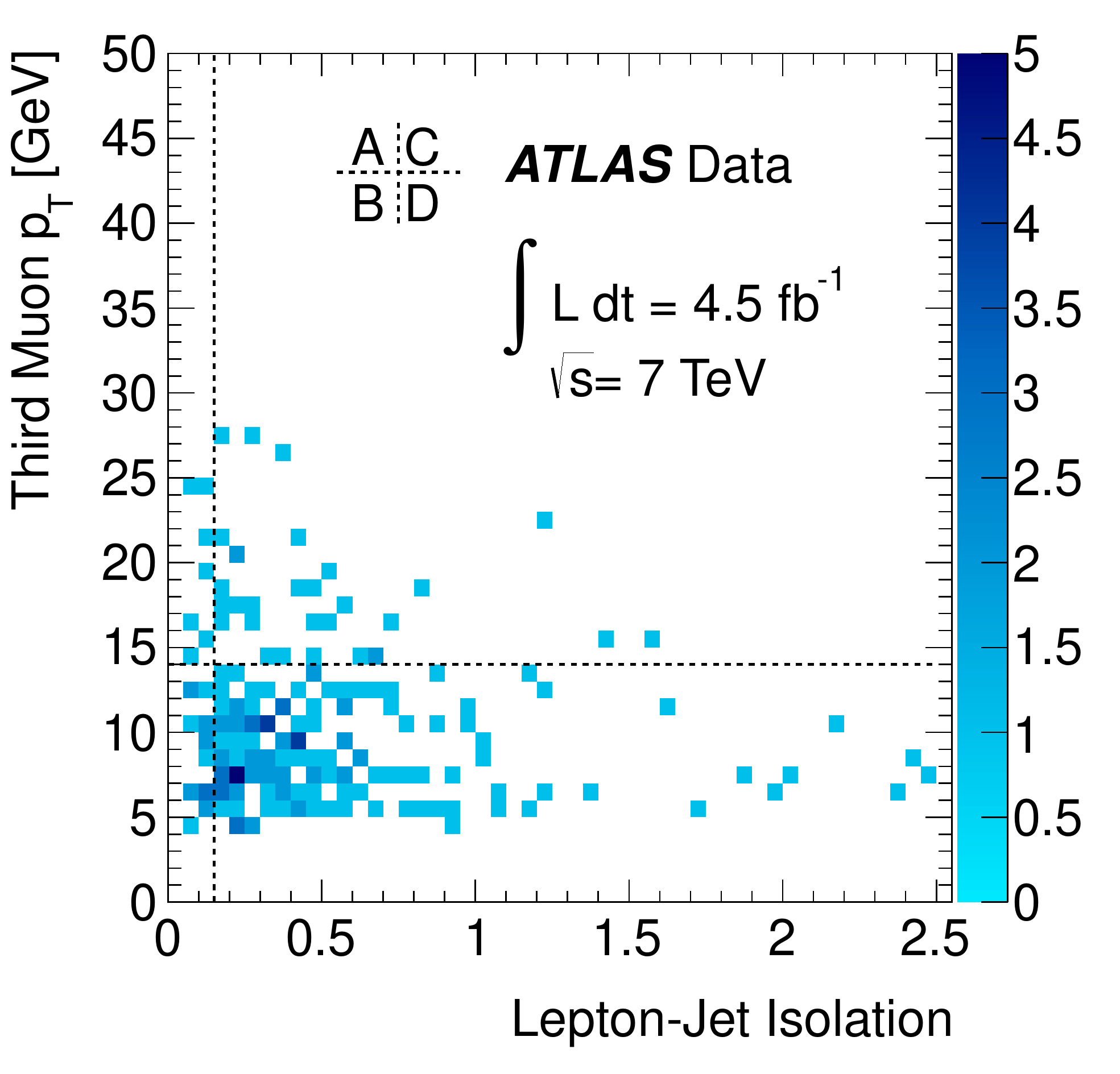}
  }
  \\
  \subfloat[][]{
    \includegraphics[width=2.05in]{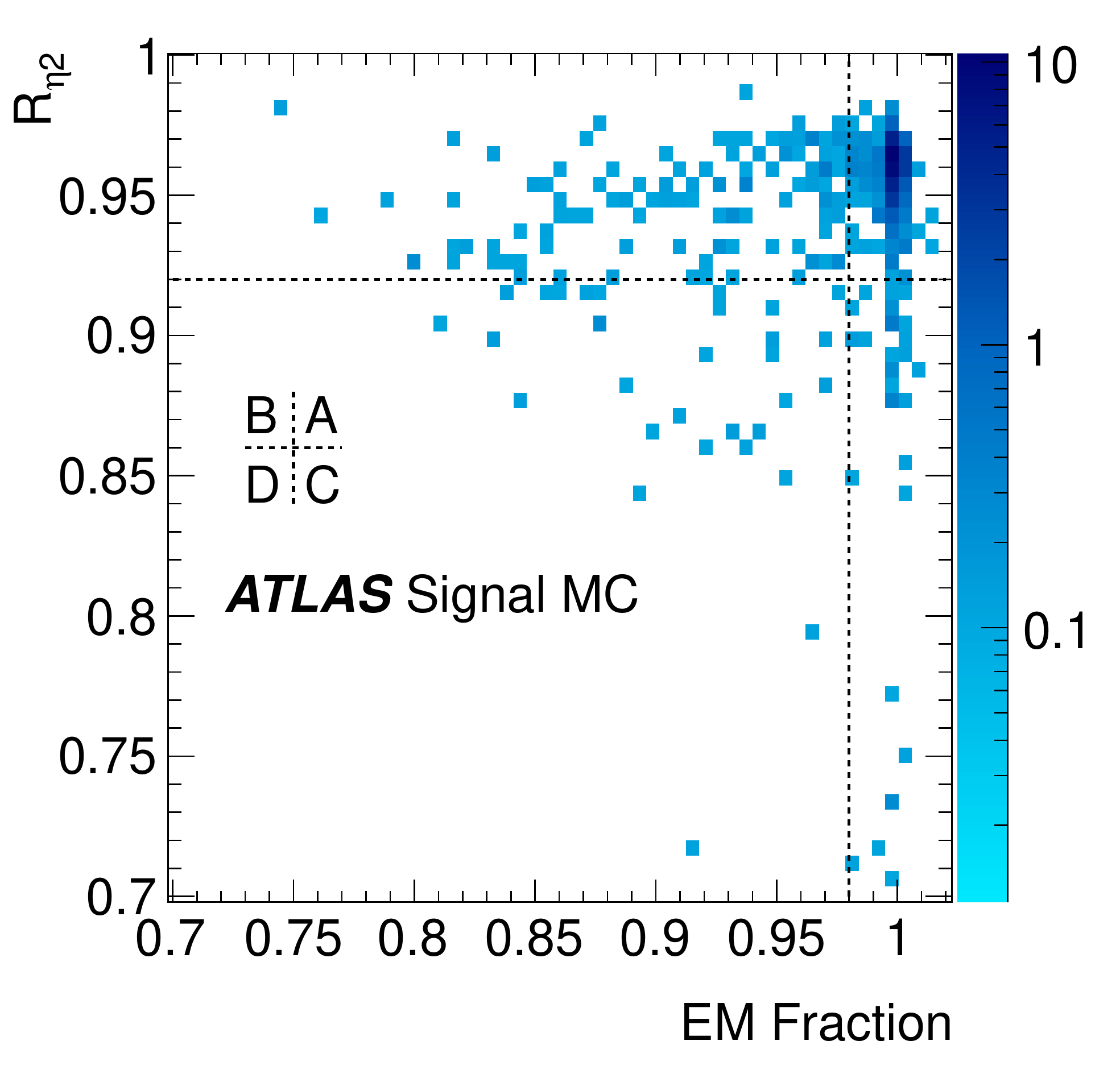}
  }
  \subfloat[][]{
   \includegraphics[width=2.05in]{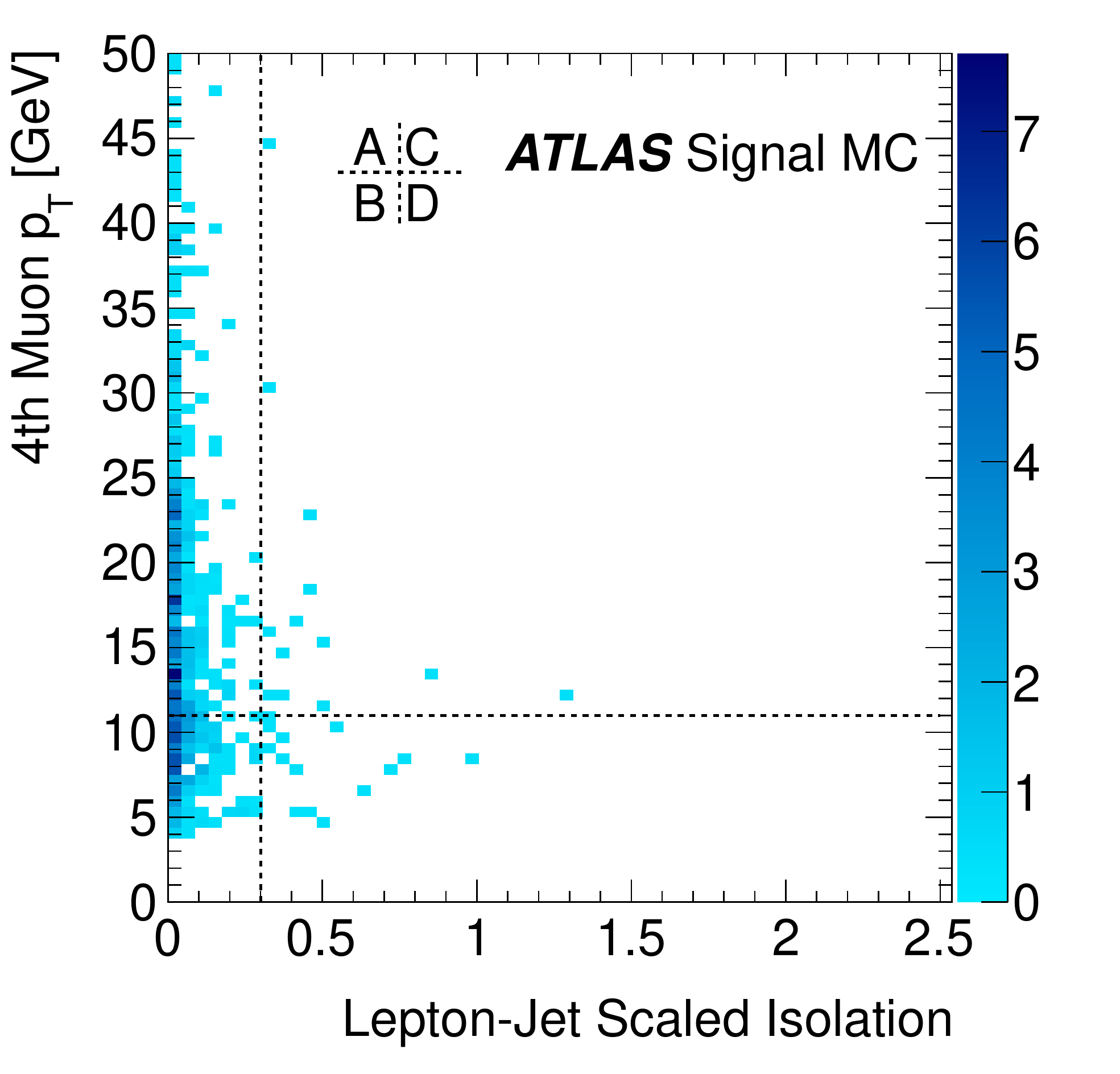}
  }
  \subfloat[][]{
   \includegraphics[width=2.05in]{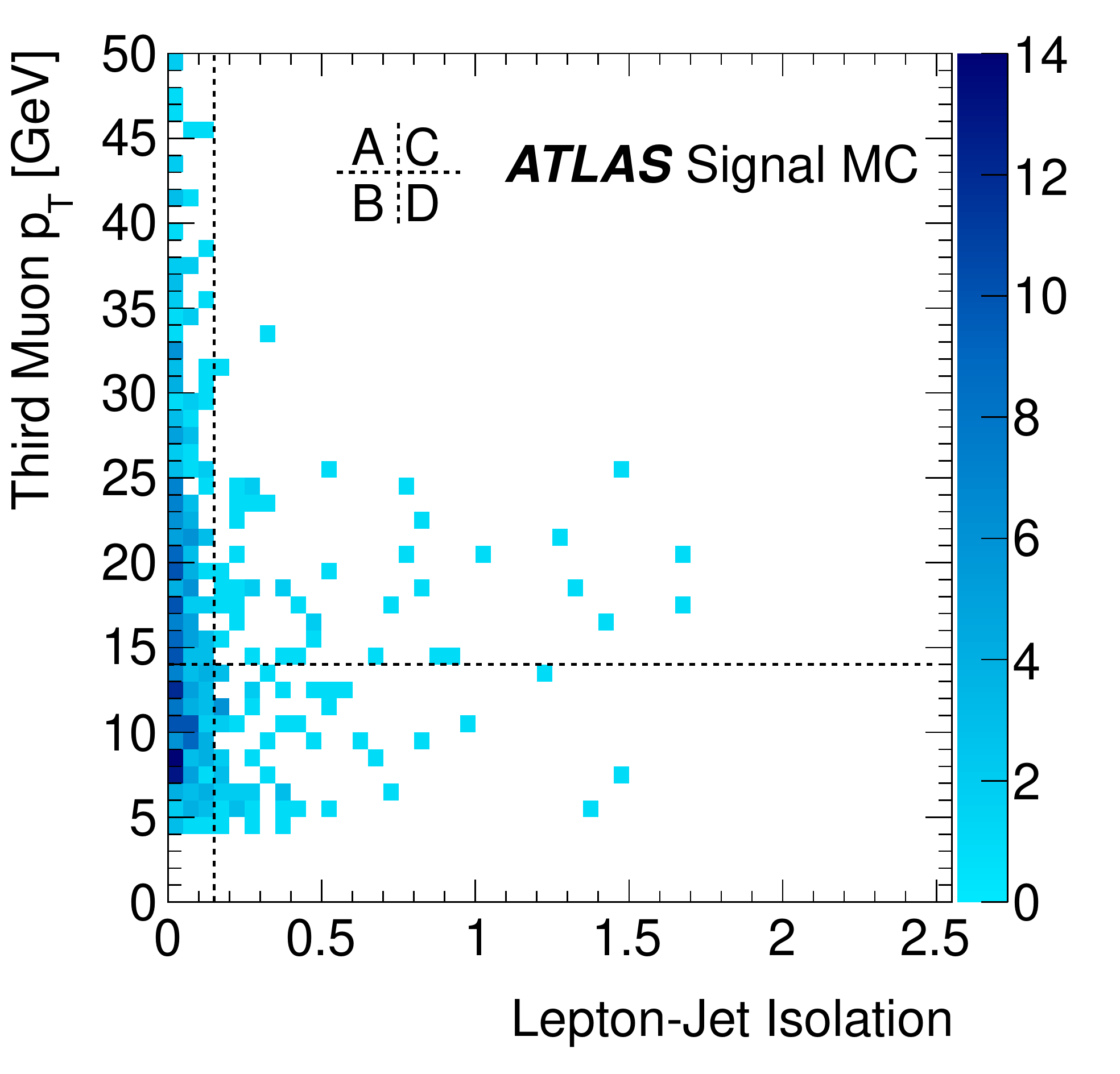}
  }
 \caption{Variables used for ABCD-likelihood method from data
   (top) and MC signal (bottom) using \alphad\ = 0.1 and \mgammad\
    = 300~MeV for the electron channel and \alphad\ = 0.0 and \mgammad\
    = 300~MeV for the two muon channels. The
   dashed black lines show the cuts used to define the four
   regions. Shown are (left) Electron-jet: $R_{\eta 2}$ vs EM fraction;
   (centre) double muon-jet: scaled isolation vs \pt\ of the
   fourth muon; (right) single muon-jet: scaled isolation vs \pt\ 
   of the third muon.
   \label{fig:ABCDplots}
 }
\end{figure*}

In the lepton \pt\ and dilepton invariant mass ranges relevant to this
study, the level of the background is best estimated using a
data-driven method, rather than by MC simulation where the number of
events is low and the backgrounds may be poorly modeled. This paper
uses an ABCD-likelihood method to determine the lepton-jet backgrounds
which was cross-checked with a tag-and-probe fake-rate estimate. The
traditional implementation of the ABCD method consists of using two
uncorrelated or loosely correlated variables from the event selection
to define four regions labeled A, B, C and D, as illustrated in 
Fig.~\ref{fig:ABCDplots}. The background in the
signal region is estimated by taking the ratio of events in the
adjacent regions. This method breaks down in the presence of
significant signal contamination in the side-band regions, or when
there are too few events. The ABCD-likelihood method addresses both of
these issues. A likelihood function, formed from the product of
Poisson probability functions describing the signal and background
expectations, is fit to all four of the regions simultaneously.

The likelihood takes the form:

\begin{equation}
  L(n_\mathrm{A}, n_\mathrm{B}, n_\mathrm{C}, n_\mathrm{D} | \mu, \theta_\mu) = \prod_{i=\mathrm{A},\mathrm{B},\mathrm{C},\mathrm{D}} \frac{e^{-\mu_i} \mu_i^{n_i}}{n_i !}
\end{equation}

where $n_{\mathrm{A}}$, $n_{\mathrm{B}}$, $n_{\mathrm{C}}$, and
$n_{\mathrm{D}}$ are the numbers of events
observed in each of the four regions, and $\mu_{\mathrm{A}}$, $\mu_{\mathrm{B}}$, $\mu_{\mathrm{C}}$,
and $\mu_{\mathrm{D}}$ are linear combinations of signal ($\mu$) and multi-jet
background ($\mu^U$) expectations. In region A, the expected number of
events $\mu_{\mathrm{A}} = \mu^U + \mu$. In region B, $\mu_{\mathrm{B}} = \mu^U \tau_{\mathrm{B}} + \mu
b$, where $\tau_{\mathrm{B}}$ is the ratio of background events expected in
region B to that in region A and $b$ gives the signal contamination in region
B. Similarly, the expected number of events in region C is expressed
as $\mu_{\mathrm{C}} = \mu^U \tau_{\mathrm{C}} + \mu c$. In region D, $\mu_{\mathrm{D}} = \mu^U \tau_{\mathrm{B}}
\tau_{\mathrm{C}} + \mu d$, such that the multi-jet background contribution is
determined using the product of the ratios. The signal contamination
coefficients are taken from MC simulation for each signal sample, while
$\mu^U$ and the $\tau_i$ values are allowed to float in a simultaneous
fit to the four data regions.

For the electron-jet analysis, the ABCD-likelihood method used
boundaries at $R_{\eta2} = 0.92$ and $f_{\mathrm{EM}}= 0.98$ on the
second highest-$\et$ lepton-jet to define these four regions. In
photon+jet events, the photon will typically deposit more energy in
the EM calorimeter than the hadronic jet. Using the subleading
cluster to estimate the background thus accounts for both the photon+jet
and multi-jet backgrounds.  The double (single) muon-jet analysis used
the scaled isolation variable and the \pt\ cut on the fourth (third)
muon in the event, associated with a muon-jet. The two-dimensional distributions with the A, B, C and D regions
are shown in Fig.~\ref{fig:ABCDplots}.  In the absence of signal, the
numbers of events predicted in region A for the single muon-jet
channel, the double muon-jet channel, and the double electron-jet
channel are $3.0 \pm 1.0$, $0.5 \pm 0.3$, and $15.2 \pm 2.7$,
respectively. The quoted errors are statistical only.

\subsection{Background estimation using jet probabilities}

Both the electron-jet and muon-jet analyses used a tag-and-probe method to
cross-check the amount of background in the signal region using
back-to-back hadronic jet pairs with $\pt > 30$ GeV and
$|\Delta\phi|>2$ but with different selection criteria.

For the electron-jet analysis, the tag was chosen by matching a jet
with $f_{\mathrm{EM}} <0.9$ to a trigger jet.  Using the
highest-\et\ electromagnetic cluster within $\Delta R = 0.4$ of the
probe jet as a seed for the electron-jet, the fake rate was extracted
from the probe jets that satisfied the electron-jet criteria, as well as
the probability for such electron-jets to pass the electron
trigger. Jet triggers with different \pt\ thresholds were used to
determine the rates over the full range of probe-jet \pt\
values. These probabilities were then used to calculate event weights
for the inclusive multi-jet MC sample to estimate the number of events which
would pass the electron trigger and electron-jet selection
requirements. This method predicted $14.55^{+0.23}_{-0.04}$ background
events after all analysis cuts were applied to the data.  The
quoted error is statistical only.

The double muon-jet analysis used two criteria to select either
light-quark or $b$-quark jets by requiring that either the tag jets
contain no muons and no b-tag, or the tag jets have a b-tag. The
probe jets were then used to determine the probability that a hadronic
jet could satisfy the muon selection criteria and the probability that
it could satisfy the muon-jet selection criteria, as a function of the
probe-jet \pt. The ratio of these two probabilities was used in
events containing three muons (of which at least two formed a muon-jet
and the third was embedded in a hadronic jet) to estimate the
background from multi-jet production, accounting for the flavour of the
hadronic jet. This method predicted $2.2 \pm 0.9$ events from multi-jet
production. The quoted error is statistical only. 

The fake rates for muon-jets and electron-jets were found to be consistent
with those obtained from the ABCD-likelihood method, which were
discussed in Section~\ref{sec:bkg_est} and are summarized in Table~\ref{tab:events}.
This cross-check thus validates the background estimates.

\section{Results and interpretation}

Table~\ref{tab:events} shows the number of events passing all analysis
cuts compared to the background expectation from the ABCD-likelihood
method. A slight excess is observed in both the single and the double
muon-jet signal regions corresponding to p-values (the probability the
background process would produce at least this many events) of 0.06
and 0.04, respectively. The acceptance times trigger, reconstruction,
and selection efficiency for the various signal points are listed in
Table~\ref{tab:acc}. It ranges from about 0.4\% to 10\% depending on
the model parameter \alphad, the mass of the dark-photon, and the
analysis channel.
The estimate of the background from the ABCD-likelihood method has a
large statistical error, which reduces the expected sensitivity of the
analysis. The systematic uncertainty on the ABCD-likelihood method
due to correlation between the variables,
3\% (4\%) for the single (double) muon-jets channel, is small by
comparison.

Table~\ref{tab:systematics} lists the systematic uncertainties on the
signal yields. The possible mismodelling of track reconstruction at
very small opening angles (``Offline $\Delta R$ Efficiency'' in
Table~\ref{tab:systematics}) introduces a $\sim$10\% systematic error
on the signal acceptance. The size of the systematic uncertainty on
the acceptance was estimated by measuring the tracking efficiency
using a tag-and-probe method with $J/\psi$ data and MC. For $\Delta R
> 0.05$ the data and MC agree to within $\sim$4\%. However, a
systematic variation of $\sim$10\% is observed in the efficiency for the smaller $\Delta
R$ region, which is probably due to a slightly softer \pt\
distribution in the MC than in the data. Systematic errors are also
assigned to the determination of the luminosity, the modelling of the
trigger acceptance, the modelling of the lepton reconstruction
efficiency, and the modelling of each of the analysis cuts.

\begin{table}[hbt]
  \footnotesize
  \centering
  \begin{tabular}{| l |c | c | c |}
    \hline
                                    & Electron LJ  & 1 Muon LJ     & 2 Muon LJ     \\ \hline
Data                                & 15           & 7             & 3             \\
All background                      & 15.2$\pm$2.7 & $3.0 \pm 1.0$ & $0.5\pm 0.3$  \\

\hline
\hline
 \end{tabular}
 \caption{The number of events in the signal region (A) observed in
   data and expected from background sources estimated with the
   ABCD-likelihood method. The quoted error is statistical only.\label{tab:events}} 
\end{table}

\begin{table}[hbt]
  \footnotesize
  \centering
  \begin{tabular}{| c c |c | c | c |}
    \hline
    \multicolumn{2}{|l|}{Signal Parameters} & Electron LJ   & 1 Muon LJ          & 2 Muon LJ            \\  
    \alphad & \mgammad [MeV] & A$\times \epsilon$ [\%] & A$\times \epsilon$ [\%] & A$\times \epsilon$ [\%] \\
    \hline
    0.0     & 150    & 3.01 $\pm$ 0.30          &                          &                          \\
    0.0     & 300    & 2.7 $\pm$ 0.5            & 4.3 $\pm$ 0.9            & 9.2 $\pm$ 0.9            \\
    0.0     & 500    & 1.8 $\pm$ 0.5            & 1.7 $\pm$ 1.3            & 8.5 $\pm$ 1.1            \\ 
    \hline
    0.10     & 150    & 2.69 $\pm$ 0.23          &                          &                          \\
    0.10     & 300    & 1.04 $\pm$ 0.19          & 3.7 $\pm$ 0.5            & 7.10 $\pm$ 0.39          \\
    0.10     & 500    & 1.17 $\pm$ 0.23          & 5.0 $\pm$ 0.8            & 8.1 $\pm$ 0.6            \\ 
    \hline
    0.30     & 150    & 2.49 $\pm$ 0.22          &                          &                          \\
    0.30     & 300    & 0.80 $\pm$ 0.13          & 2.16 $\pm$ 0.29          & 7.47 $\pm$ 0.42          \\
    0.30     & 500    & 0.37 $\pm$ 0.10          & 3.16 $\pm$ 0.46          & 6.23 $\pm$ 0.43          \\
\hline
\hline
 \end{tabular}
 \caption{The acceptance times trigger, reconstruction, and selection efficiency (A$\times \epsilon$)
   expected in the signal region (A) from various signal hypotheses
   for the three different lepton-jet (LJ) channels. Note that the
   \mgammad\ = 150 MeV dark photon cannot decay to muons.\label{tab:acc}}
\end{table}

\begin{table*}[ht]
  \footnotesize
  \centering
\begin{tabular}{|c|c|c|c|}
\hline
                              & Electron LJ [\%] & 1 Muon LJ [\%] & 2 Muon LJ [\%] \\ \hline
Luminosity                    & 3.9         & 3.9     & 3.9     \\ \hline
Trigger Efficiency            & 1.5         & 2.0     & 3.6     \\ \hline
Offline $\Delta R$ Efficiency & 13.0        & 10.7    & 10.7    \\ \hline
Lepton Momentum Scale                  &   0.6       & 1.0     & 1.0     \\ \hline
Isolation                     &   5.2       &  $<0.1$ & $<0.1$  \\ \hline
$R_{\eta 2}$ and $w_{\eta 2}$ Efficiency &   8.0       & NA      & NA      \\ \hline
$\fHT$ Efficiency            &   1.0       & NA      & NA      \\ \hline
$\fEM$ Efficiency            &   3.0       & NA      & NA      \\ \hline
Muon Momentum Resolution               & NA          & $< 1.0$ & $< 1.0$ \\ \hline
\end{tabular}
\caption{Contributions to the systematic uncertainty on the signal yields for the three
different lepton-jet (LJ) channels given as percentages. A ``NA''
means this source does not apply.}
\label{tab:systematics}
\end{table*}

The 95\% confidence-level upper limits on the number of expected
events from new phenomena producing collimated pairs of prompt leptons
were calculated using the CLs method~\cite{CLs} with a log-likelihood
ratio (LLR) test statistic. Ensembles of pseudo-experiments were
generated for the signal-only hypothesis and the signal+background
hypothesis, varying the LLR according to the statistical and
systematic uncertainties. The upper limits were determined by
performing a scan of $p$-values corresponding to LLR values larger
than the one observed in data.
For broad applicability, the limits are expressed in terms of the
signal cross section times branching ratio to the final state under
consideration, using the expected signal acceptance for each of the
pairings of dark photon masses and dark sector gauge coupling
parameter values in Table~\ref{tab:limits}.

\begin{table}[hbt]
  \footnotesize
  \centering
  \begin{tabular}{ | c  c | c | c | c | }
    \hline
\multicolumn{2}{|l|}{Signal Parameters} & Electron LJ   & 1 Muon LJ          & 2 Muon LJ \\  
\alphad & \mgammad [MeV] & Obs   (Exp) pb & Obs   (Exp) pb & Obs    (Exp)  pb           \\
    \hline
0.0     & 150            & 0.082 (0.082)  & -              & -                          \\
0.0     & 300            & 0.11  (0.11)   & 0.060 (0.035)  & 0.017  (0.011)             \\
0.0     & 500            & 0.20  (0.21)   & 0.15  (0.090)  & 0.019  (0.012)             \\ \hline
0.10    & 150            & 0.096 (0.10)   & -              & -                          \\
0.10    & 300            & 0.37  (0.37)   & 0.064 (0.036)  & 0.018  (0.011)             \\
0.10    & 500            & 0.39  (0.39)   & 0.053 (0.035)  & 0.018  (0.011)             \\ \hline
0.30    & 150            & 0.11  (0.11)   & -              & -                          \\
0.30    & 300            & 0.40  (0.40)   & 0.099 (0.055)  & 0.020  (0.012)             \\
0.30    & 500            & 1.2   (1.2)    & 0.066 (0.043)  & 0.022  (0.015)             \\

\hline
\hline
 \end{tabular}
  \caption{Observed (Expected) upper limits for cross section times
    branching ratio ($\sigma \times$ BR) to the final state under
    consideration, in units of pb for the three different lepton-jet
    (LJ) channels.
\label{tab:limits}}
\end{table}

The observed limits in the electron-jets channel are in good agreement
with the expected limits, and are the first inclusive study of prompt
electron-jets at the LHC. The limits in the muon-jet channels are slightly
higher than expected as a result of the slight excesses, but are
within $2 \sigma$ of the SM expectation for both channels. The limits on the production of prompt
lepton-jets in the muon-jet channel improve upon previous results by an
order of magnitude.

\section{Conclusions}
A search for collimated pairs of muons or electrons, lepton-jets, has
been performed using nearly 5~fb$^{-1}$ of $pp$ collisions at
$\sqrt{s}=7$~TeV recorded with the ATLAS detector at the LHC. Such final states
have been proposed as a possible explanation of recently observed
anomalies in cosmic-ray and dark matter direct-detection
experiments. No significant excess of data compared to the SM expectation was
observed in any of the three channels, and 95\% confidence-level upper
limits have been computed on the cross section times branching ratio
for several parameter values of a Hidden Valley model. The limits
range from 0.017 to 1.2 pb.

\mbox{ }\\


\section{Acknowledgements}

We thank CERN for the very successful operation of the LHC, as well as the
support staff from our institutions without whom ATLAS could not be
operated efficiently.

We acknowledge the support of ANPCyT, Argentina; YerPhI, Armenia; ARC,
Australia; BMWF and FWF, Austria; ANAS, Azerbaijan; SSTC, Belarus; CNPq and FAPESP,
Brazil; NSERC, NRC and CFI, Canada; CERN; CONICYT, Chile; CAS, MOST and NSFC,
China; COLCIENCIAS, Colombia; MSMT CR, MPO CR and VSC CR, Czech Republic;
DNRF, DNSRC and Lundbeck Foundation, Denmark; EPLANET, ERC and NSRF, European Union;
IN2P3-CNRS, CEA-DSM/IRFU, France; GNSF, Georgia; BMBF, DFG, HGF, MPG and AvH
Foundation, Germany; GSRT and NSRF, Greece; ISF, MINERVA, GIF, DIP and Benoziyo Center,
Israel; INFN, Italy; MEXT and JSPS, Japan; CNRST, Morocco; FOM and NWO,
Netherlands; BRF and RCN, Norway; MNiSW, Poland; GRICES and FCT, Portugal; MERYS
(MECTS), Romania; MES of Russia and ROSATOM, Russian Federation; JINR; MSTD,
Serbia; MSSR, Slovakia; ARRS and MVZT, Slovenia; DST/NRF, South Africa;
MICINN, Spain; SRC and Wallenberg Foundation, Sweden; SER, SNSF and Cantons of
Bern and Geneva, Switzerland; NSC, Taiwan; TAEK, Turkey; STFC, the Royal
Society and Leverhulme Trust, United Kingdom; DOE and NSF, United States of
America.

The crucial computing support from all WLCG partners is acknowledged
gratefully, in particular from CERN and the ATLAS Tier-1 facilities at
TRIUMF (Canada), NDGF (Denmark, Norway, Sweden), CC-IN2P3 (France),
KIT/GridKA (Germany), INFN-CNAF (Italy), NL-T1 (Netherlands), PIC (Spain),
ASGC (Taiwan), RAL (UK) and BNL (USA) and in the Tier-2 facilities
worldwide.

\bibliographystyle{elsarticle-num}

\bibliography{bibliography}

\clearpage
\begin{flushleft}
{\Large The ATLAS Collaboration}

\bigskip

G.~Aad$^{\rm 48}$,
T.~Abajyan$^{\rm 21}$,
B.~Abbott$^{\rm 111}$,
J.~Abdallah$^{\rm 12}$,
S.~Abdel~Khalek$^{\rm 115}$,
A.A.~Abdelalim$^{\rm 49}$,
O.~Abdinov$^{\rm 11}$,
R.~Aben$^{\rm 105}$,
B.~Abi$^{\rm 112}$,
M.~Abolins$^{\rm 88}$
% ...
\bigskip
\\
% ...
$^{11}$ Institute of Physics, Azerbaijan Academy of Sciences, Baku, Azerbaijan\\
$^{12}$ Institut de F{\'\i}sica d'Altes Energies and Departament de F{\'\i}sica de la Universitat Aut{\`o}noma de Barcelona and ICREA, Barcelona, Spain\\
% ...
$^{21}$ Physikalisches Institut, University of Bonn, Bonn, Germany\\
% ...
$^{48}$ Fakult{\"a}t f{\"u}r Mathematik und Physik, Albert-Ludwigs-Universit{\"a}t, Freiburg, Germany\\
$^{49}$ Section de Physique, Universit{\'e} de Gen{\`e}ve, Geneva, Switzerland\\
% ...
$^{88}$ Department of Physics and Astronomy, Michigan State University, East Lansing MI, United States of America\\
% ...
$^{105}$ Nikhef National Institute for Subatomic Physics and University of Amsterdam, Amsterdam, Netherlands\\
% ...
$^{111}$ Homer L. Dodge Department of Physics and Astronomy, University of Oklahoma, Norman OK, United States of America\\
$^{112}$ Department of Physics, Oklahoma State University, Stillwater OK, United States of America\\
% ...
$^{115}$ LAL, Universit{\'e} Paris-Sud and CNRS/IN2P3, Orsay, France\\
% ...
\end{flushleft}

\end{document}